\documentclass[reprint,aip,apl,superscriptaddress,showpacs,floatfix]{revtex4-1}
\usepackage[pdfborder=0 0 0]{hyperref}
\usepackage[dvips]{graphicx}
\usepackage{amsmath}
\usepackage{color}
\usepackage{amsfonts}
\usepackage{amssymb}
\usepackage{epstopdf}
\usepackage{colortbl}
\DeclareGraphicsExtensions{.eps,.png,.pdf}
\begin{document}
\title{Plasma density and ion energy control via driving frequency and applied voltage in a low pressure capacitively coupled plasma discharge}
\author{Sarveshwar Sharma} 
\email[e-mail: ]{sarvesh@ipr.res.in}
\affiliation{Institute for Plasma Research, Bhat, Gandhinagar 382 428, India}
\affiliation{Homi Bhabha National Institute, Anushaktinagar, Mumbai 400 094, India}
 \author{Abhijit Sen}
 \affiliation{Institute for Plasma Research, Bhat, Gandhinagar 382 428, India}
\author{N. Sirse}
\author{M. M. Turner}
\author{A. R. Ellingboe}
\affiliation{School of Physical Sciences and National Centre for Plasma Science and Technology (NCPST),
Dublin City University, Dublin 9, Republic of Ireland}

\newcommand{\beq}{\begin{equation}}
\newcommand{\eeq}{\end{equation}}
\newcommand{\beqstar}{\[}
\newcommand{\eeqstar}{\]}
\newcommand{\bea}{\begin{eqnarray}}
\newcommand{\eea}{\end{eqnarray}}
\newcommand{\beastar}{\begin{eqnarray*}}
\newcommand{\eeastar}{\end{eqnarray*}}
\begin{abstract} 
\noindent The dynamical characteristics of a single frequency low pressure capacitively coupled plasma (CCP) device  under varying applied RF voltages and driving frequencies are studied using particle-in-cell/Monte Carlo collision simulations. An operational regime is identified where for a given voltage the plasma density is found to remain constant over a range of driving frequencies and to then increase rapidly as a function of the driving frequency. The threshold frequency for this mode transition as well as the value of the constant density is found to increase with an increase in the applied voltage. Over the constant density range, for a given voltage, the sheath width is seen to increase as a function of the increasing driving frequency, thereby changing the ion energy without affecting the ion density. Our parametric study thus indicates that the twin knobs of the applied voltage and driving  frequency offer a means of independently controlling the density and the ion energy in a low pressure CCP device that may be usefully exploited for plasma processing applications.
\end{abstract}

\maketitle

Single frequency capacitively coupled (CCP) plasma devices have been extensively employed in the plasma processing industry for applications like the etching and deposition of thin films\cite{Lib_Wiley_2005}. A CCP device consists of two parallel electrodes between which a plasma discharge is struck by biasing the electrodes with a radio frequency power supply that is typically operated at a single frequency of $13.56$ MHz. The bulk plasma is separated from the electrodes by the formation of space charge sheaths whose thicknesses oscillate at the driving frequency. The sheath and bulk plasma properties of the discharge are influenced by a number of factors such as the driving frequency, the applied voltage, the background neutral pressure, the gap between the electrodes and the relative areas of the electrodes. For material processing applications the plasma density and the ion energy are important parameters that have a significant impact on the quality and rate of etching/deposition processes. In low pressure discharges the ions gain energy mainly through their acceleration from the DC bias field within the sheath while the electrons gain energy through a stochastic heating process through their interaction with the high-voltage oscillating sheath \cite{Lib_16_1988, Lib_26_1998, God_16_1972, Pop_57_1985, Kag_89_2002, Turner_75_1995, Kawa_13_2006, Sharma_2013, Goz_87_2001, Sharma_22_2013, Sharma_21_2014, Kag_34_2006, Sharma_55_2015, Sharma_20_2013, Sur_19_1991, Turner_42_2009}. The plasma density is controlled both by the applied voltage and the driving frequency. Typically it increases linearly with the applied voltage and quadratically with the driving frequency. In the traditional single frequency CCP device the plasma density and the ion energy cannot be changed independently since an increase in the RF power changes both the density of the plasma and the sheath voltage which affects the energy of the ions\cite{Lib_Wiley_2005}. Such a limitation of the single frequency CCP has led to searches for alternative configurations as well as explorations of different parametric regimes. The dual frequency CCP has been a major development in this direction and is now quite widely used in the industry\cite{Kawa_13_2006, Goto_1993_6_58, Perret_2005_86_021501, Sharma_46_2013, Boyle_2004_37_697}.  Plasma excitation by non-sinusoidal “tailored” waveforms \cite{Sharma_PSST_24_2015} has also been  proposed for the independent control of energy and flux but is difficult to implement in an industrial grade reactor due to challenges in designing the impedance matching network \cite{Traver_APL_101_2012}. The operational regime of the single frequency CCP has also been studied for much higher frequencies, up-to several tens of MHz, with the objective of attaining enhanced plasma processing rates with lesser damage to the substrate over a large wafer area \cite{How_10_1992, Suren_59_1991}. Some of these studies have revealed interesting basic findings about the nature of the discharge behavior at higher frequencies. For example, in contrast to the conventional view, the plasma density has been found not to have a quadratic dependence on the driving frequency but to suffer an abrupt step like jump at a distinct driving frequency \cite{Wil_24_2015}. The step like behavior which was found at neutral pressures of 1.1 Pa was seen to smooth out as the pressure was increased and to revert to a quadratic dependence on the driving  frequency at higher pressures. The low pressure regime of operation thus has distinctly different characteristics compared to the standard CCP particularly when operated at high frequencies. The effect of  driving frequency variation on the plasma density was also studied in \cite{Sharma_23_2016} 
at a lower pressure of $5$ mTorr and it was noted that the plasma density maintained a constant value over a range of driving frequencies upto a critical frequency value before rising with the driving frequency. In this work we carry out a more detailed investigation of this regime and study the extent and nature of this constant density region as a function of the driving frequency as well as the applied voltage. We find that the extent of the constant density regime is a function of the applied voltage i.e. the threshold frequency after which the density starts rising increases with an increase in the applied voltage. We further find that over the constant density regime, for a given voltage, the sheath width at the electrode increases with the increase in the driving frequency leading to a concomitant increase  in the ion energy without an accompanying rise in density. When the applied voltage is increased for a fixed driving frequency lying within the constant density regime, the density value shifts to a higher magnitude thereby increasing the ion flux without affecting the ion energy (since the sheath width remains the same). Our simulations thus reveal an interesting possibility of independent control of the ion density and ion energy by an optimal choice of the external driving frequency and applied voltage values - a finding that could be potentially useful for plasma processing applications.

\begin{figure}[t]
\centering
\includegraphics[width=0.5\textwidth]{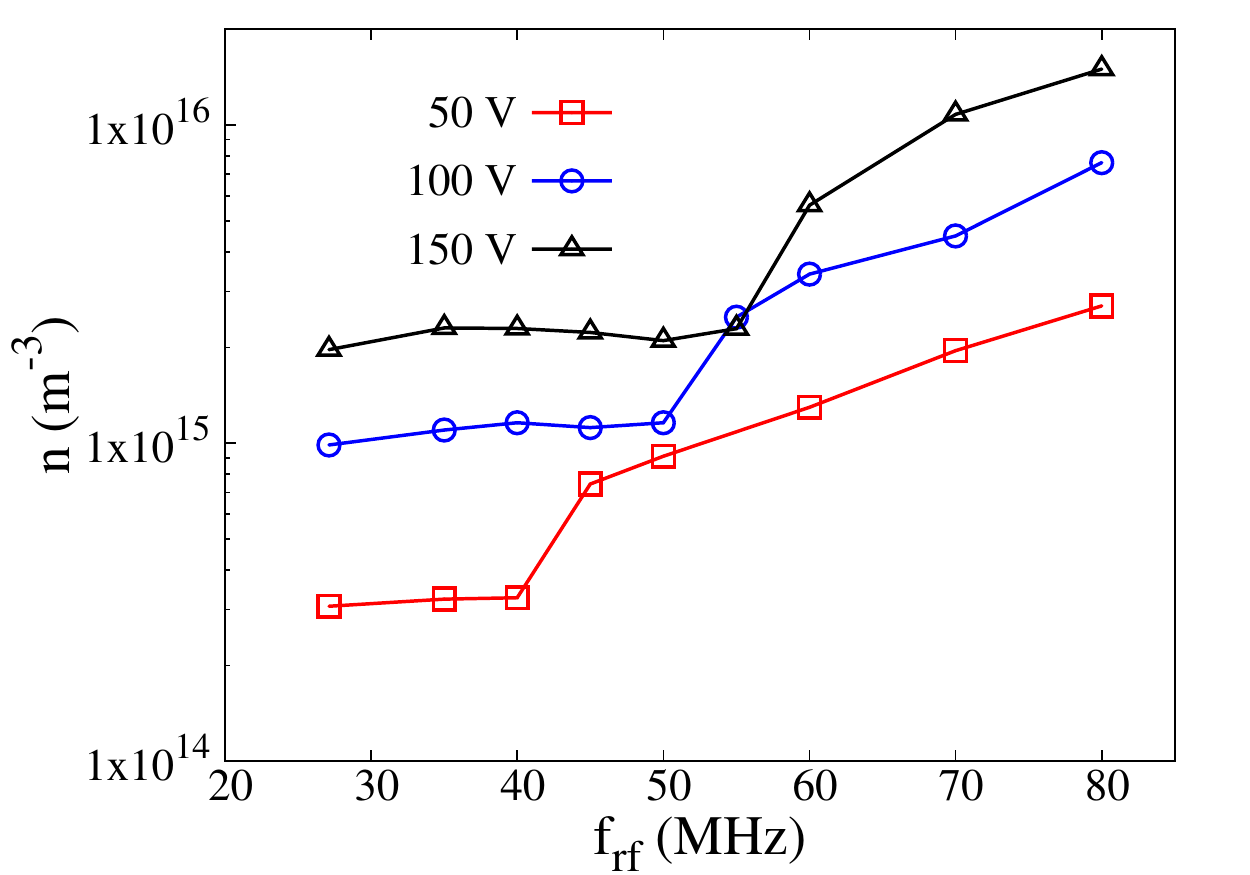}
\caption{Plasma Density at the center of discharge as a function of the applied voltage and RF frequency.}
\label{fig:Fig_1}
\end{figure}

\noindent Our simulations have been carried out with a $1D3V$ electrostatic particle-in-cell/Monte Carlo collisions (PIC/MCC) code that was developed by the group at Dublin City University, Ireland \cite{Turner_22_2013}. The code has been extensively tested and successfully employed in a number of past investigations of the dynamics of CCP plasmas \cite{Turner_22_2013, Conway_22_2013, Boyle_13_2004, Taroni_37_2004, Turner_76_1996, Sharma_24_2017, Sharma_23_2016, Turner_20_2013}. The code takes account of 
all important particle interactions like electron-neutral (elastic, inelastic and ionization) and ion-neutral (elastic, inelastic and charge exchange) collisions but omits higher order processes like multi-step ionization, metastable pooling, partial de-excitation, super elastic collisions etc. The simulations have been carried out for an Argon plasma where
the production of metastables (\textit{i.e.} Ar*, Ar**) is also included \cite{Sharma_POP_25_2018}. An appropriate choice of spatial step size (\textit{i.e.} smaller than the Debye length) and a temporal step size (\textit{i.e.} less than the electron plasma frequency) are taken to satisfy the numerical accuracy and stability criteria of the PIC simulations\cite{Bird_1991, Hock_1988}. We assume that the electrodes are planar and parallel to each other with infinite dimension so that a one dimensional simulation is appropriate. The model CCP can be operated either in the current or in the voltage driven modes. For our present analysis we have  selected the voltage driven mode of operation. Both electrodes are perfectly absorbing for electrons and ions and secondary electron emission is ignored. The gap between the electrodes is taken to be $3.2$ cm and the simulation region is divided into a grid consisting of $512$ cells with each cell being populated by 100 particles. The background neutral gas is distributed uniformly at a temperature of $300$ K,  and the initial temperature of ions is similar to that of neutrals. One of the electrodes is grounded while the other is driven by an RF voltage having the following waveform,
\begin{equation}
V_{rf}(t) = V_0sin(2\pi f_{rf}t+\phi)
\label{equation1}
\end{equation}
A schematic diagram of the model CCP discharge is shown in Ref.\cite{Lib_Wiley_2005,Sharma_24_2017}. The amplitude of the applied voltage is varied from $50$ V to $150$ V and the driving frequency is varied from $27.12-80$ MHz while the pressure is held constant at $5$ mTorr.

In our simulations we have primarily focused our attention on operational regimes where the plasma peak density remains independent of the variation of the driving frequency over a certain range \cite{Sharma_23_2016,Wil_24_2015}. For a given value of applied voltage we have made a number of simulation runs to determine the stationary state values of various plasma quantities like the density, the effective electron temperature (i.e. $T_{eff}$), the sheath width and ion energy for different values of the driving frequency. Here $T_{eff}$ is defined as
\begin{equation}
T_{eff} = \left(\frac{2}{3} \right) \frac{\int \varepsilon F(\varepsilon) d\varepsilon}{\int F(\varepsilon) d\varepsilon} 
\label{equation3}
\end{equation}
Where $F (\varepsilon)$ is the self–consistent EEDF and $\varepsilon$ is the electron energy obtained from the simulation. The runs are then repeated for a different constant value of the applied voltage. Fig.~(\ref{fig:Fig_1}) provides a composite plot of the plasma density at the center of discharge  
over a range of driving frequencies for three different values of the applied voltage, namely, $50$ V, $100$ V and $150$ V.  

\begin{figure*}
\centering
\includegraphics[width=0.82\textwidth]{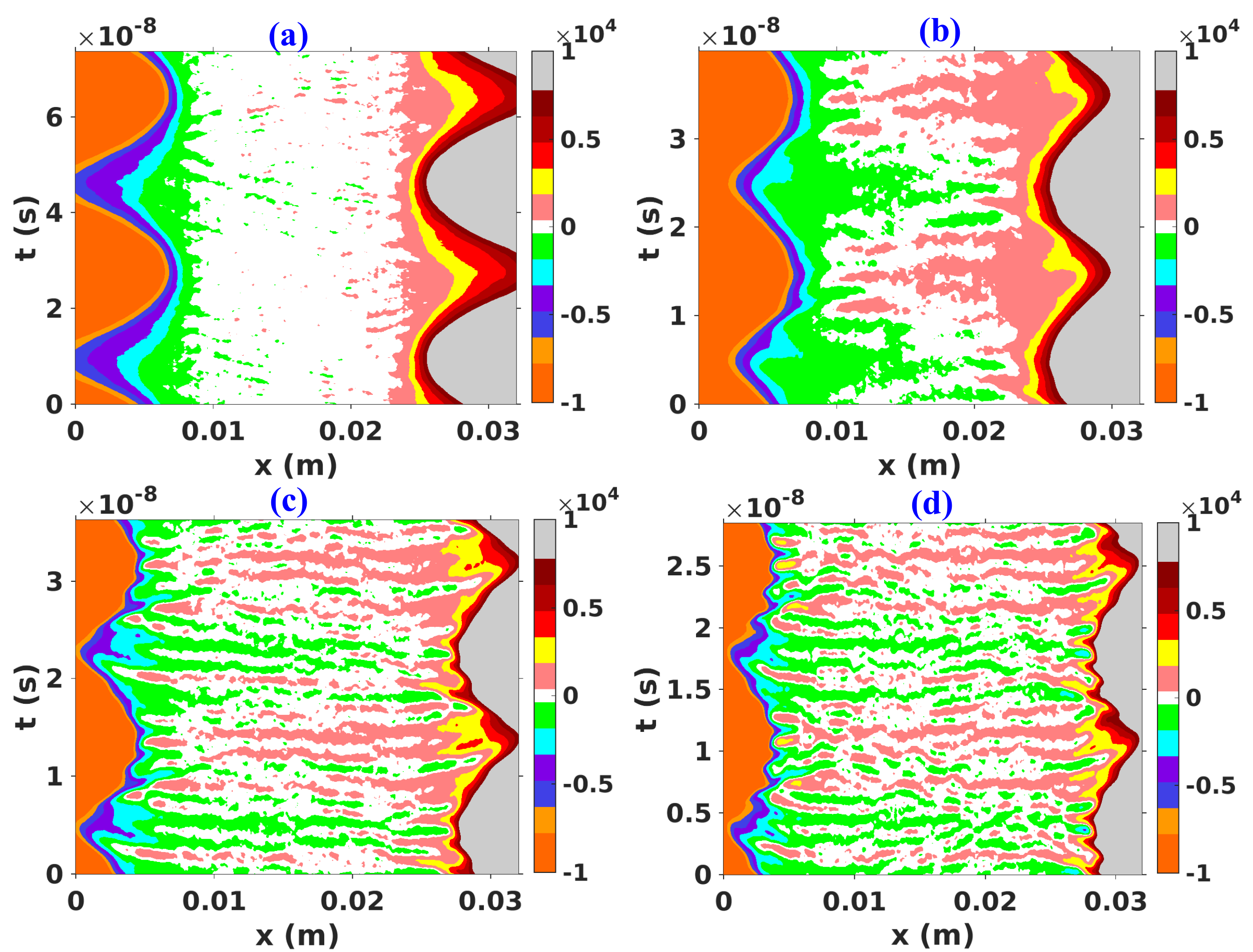}
\caption{Spatio-temporal evolution of the electric field at a) $27.12$ MHz b) $50$ MHz c) $55$ MHz and d) $70$ MHz driving frequency for a constant applied voltage of $100$ V. (Reproduced with permission from Phys. Plasmas 23(11), 110701 (2016) \cite{Sharma_23_2016}).}
\label{fig:Fig_2}
\end{figure*}

The figure shows that at a particular applied voltage there is a range of driving frequencies over which the plasma density is nearly constant e.g. at $50$ V in the range of $27.12-40$ MHz (rise in density at $45$ MHz), at $100$ V from $27.12-50$ MHz (rise in density at $55$ MHz) and at $150$ V from $27.12-55$ MHz (rise in density at $60$ MHz) respectively. The constant density values at $50$ V, $100$ V and $150$ V are $\sim$ $3.2\times10^{14}$ $m^{-3}$ (from $27.12-40$ MHz), $\sim$ $1\times10^{15}$ $m^{-3}$ (from $27.12-50$ MHz) and $\sim$ $2.2\times10^{15}$ $m^{-3}$ (from $27.12-55$ MHz) respectively. 
The corresponding ion flux also remains nearly constant in this driving frequency range. The typical values of ion flux are $5-7\times10^{17}$ $m^{-2}s^{-1}$ ($27.12-40$ MHz, $1.2-1.8\times10^{18}$ $m^{-2}s^{-1}$ ($27.12-50$ MHz), $2.5-3.5\times10^{18}$ $m^{-2}s^{-1}$ ($27.12-55$ MHz) at $50$ V, $100$ V and $150$ V respectively. 
The figure further shows that by increasing or decreasing the discharge voltage, the transition frequency after which the plasma density increases could shift towards a higher or lower side of the driving frequency spectrum respectively. 
The transition frequency in each case marks a distinct change in the behavior of the plasma response
with respect to the driving frequency. 
Such a behavior of the plasma density can be understood in terms of the nature of the transient electric field structures \cite{Sharma_23_2016} emanating from the sheath regions near to the electrodes These transients which are mainly localized near the sheath at $27.12$ MHz begin to penetrate deeper into the bulk plasma as the driving frequency is raised to $70$ MHz as can be seen from Fig.~(\ref{fig:Fig_2}). 

\begin{figure}[t]
\centering
\includegraphics[width=0.5\textwidth]{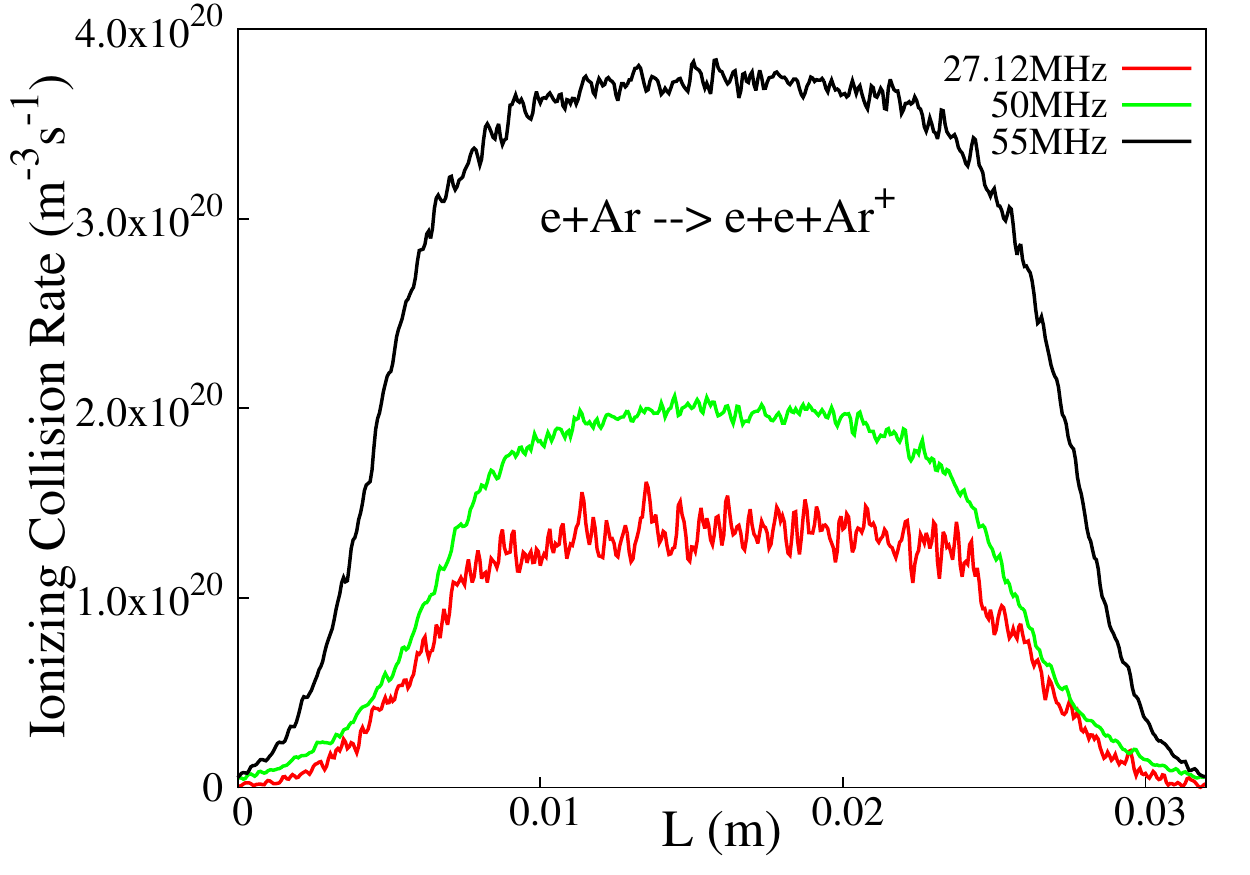}
\caption{Ionization rate profiles for $27.12$ MHz, $50$ MHz and $55$ MHz for a constant applied voltage of $100$ V.}
\label{fig:Fig_3}
\end{figure}

Below the transition frequency these transients transfer energy to the electrons in a variety of ways ranging from linear wave excitations resulting in phase mixing (at low applied voltages) to nonlinear processes that bring about a change in the electron energy distribution function (EEDF)\cite{Sharma_23_2016}. The transients thereby bring about an energy redistribution by transferring energy from high energy electrons to the bulk electrons which concomitantly influence the various collision rates in the plasma. In particular, for driving frequencies below the transition frequency, the inelastic collision rates experience a significant enhancement whereas the ionizing collision rates do not increase significantly. For example, at $100$ V the inelastic collision rate increases from $1.26\times10^{20}$ $m^{-3}s^{-1}$ (at $27.12$ MHz) to $2.2\times10^{20}$ $m^{-3}s^{-1}$ (at $50$ MHz), whereas, the ionizing collisional rate changes from $\sim$ $1.5\times10^{20}$ $m^{-3}s^{-1}$ (at $27.12$ MHz) to $\sim$ $1.9\times10^{20}$ $m^{-3}s^{-1}$ (at $50$ MHz). Thus there is a $\sim$ $75\%$ growth in the inelastic collision rate while the change in the ionizing collisional rate is about $\sim$ $27\%$. This accounts for the lack of rise in the plasma density value as a function of the driving frequency and leads to the observed near constancy of the density value below the transition frequency. Above the transition frequency both the inelastic and ionizing collision rates rise rapidly with the rise in driving frequency and explain the rise of plasma density shown in Fig.~(\ref{fig:Fig_1}). The dramatic change in the ionization rate is illustrated in Fig.~(\ref{fig:Fig_3}) where the spatial distribution of ionization rates are shown for different values of the driving frequency at a constant value of the applied voltage. As can be seen the magnitude of ionization in the central portion of the discharge jumps up significantly for $55$ MHz compared to the value at $50$ MHz for the $100$ V case.

We next turn to the energetics of the ions as a function of the driving frequency. As mentioned before, the transfer of RF energy to the ions mainly takes place in the sheath region where they get accelerated by the votage drop across the sheath and the gain in energy is a function of the maximum width of the sheath. In our simulations we have closely investigated the changes in the maximum widths of the sheaths \cite{Lib_Wiley_2005} and the results are listed in Table~\ref{Table1}. 

\begin{table}
\caption{Sheath Widths}
\centering
\begin{tabular}{c c c c c c}
\hline\hline
Voltage  &  27.12 MHz   &   35 MHz   &  40 MHz     &   50 MHz   &   55 MHz \\ [0.5ex]
\hline
50V      &  0.91 cm     &   0.98 cm  &  1.1 cm     &    -       &   -       \\
100V     &  0.79 cm     &    0.82 cm &   0.88 cm      &    1.0 cm &   -   \\ 
150V     &  0.74 cm     &    0.79 cm  &   0.86 cm      &    0.91 cm      &   0.96 cm  \\ [1ex]
\hline
\end{tabular}
\label{Table1}
\end{table}

\begin{figure}
\centering
\includegraphics[width=0.45\textwidth]{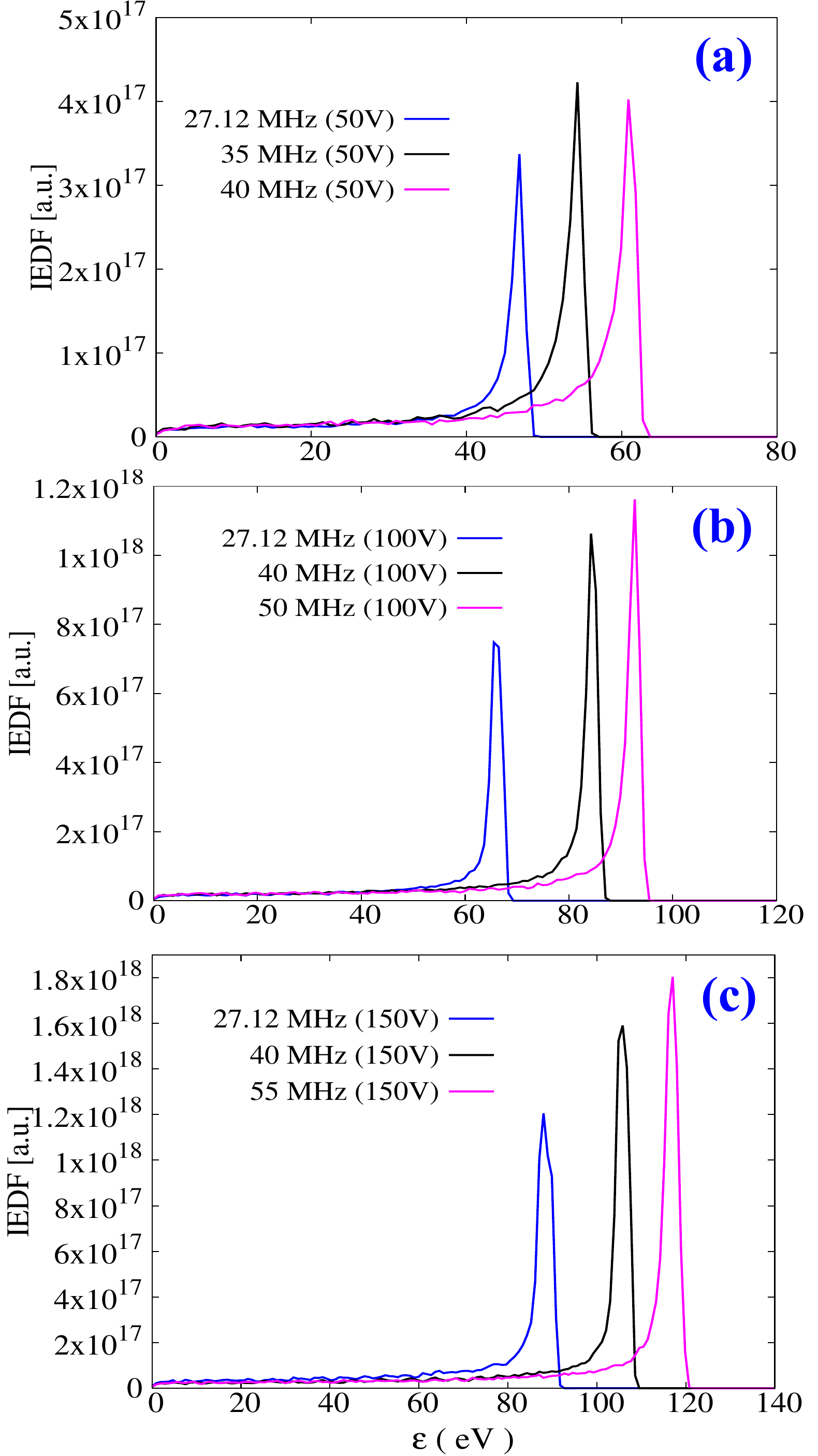}
\caption{Ion Energy Distribution functions arriving at electrode for different driving frequencies i.e. $27.12-40$ MHz, $27.12-50$ MHz and $27.12-55$ MHz  for an applied voltages of $50$ V, $100$ V and $150$ V respectively.}
\label{fig:Fig_4}
\end{figure}

As can be seen the sheath width increases systematically over the range of driving frequencies for which the density remains constant in each of the constant voltage runs. The corresponding gain in the ion energies are shown in Fig.~(\ref{fig:Fig_4}) in terms of the changes in the ion energy distribution functions for the various cases. A single energy peak is observed in all cases which corresponds to a collisionless acceleration of ions in the sheath. Furthermore, the ion transit time is about $20$ to $60$ times larger than the rf period which results in a single energy peak due to the ion response to the time-averaged voltage drop within the sheath. It is further observed that by increasing the applied voltage a broader range of ion energy control could be covered. For $50$ V, the energy range is $13$ eV (i.e. $47-60$ eV) which increases to $25$ eV (i.e. $67-92$ eV) at $100$ V and finally $32$ eV (i.e. $86-118$ eV) at $150$ V.

\noindent 
In conclusion, we have carried out a detailed investigation of the dynamical characteristics of a capacitively coupled plasma discharge when it is operated at a low pressure ($5$ mTorr) and over a range of high driving frequencies 
and variable voltages. Our PIC simulation results, for an Argon plasma, reveal the existence of operating regimes
where the plasma density (i.e. the density at the center of the discharge) can be kept constant over a range of driving frequencies while holding the applied voltage constant.
Over this range the ion energy keeps increasing with driving frequency as a result of the change in the sheath width. By raising the value of the applied voltage at a given driving frequency in the constant density range one can controllably move to a higher value
of the plasma density without changing the ion energy. Thus by manipulating the driving frequency and applied voltage it seems possible to independently control the ion energy and ion density in the discharge. Such a flexibility is contingent upon the existence of an operating regime where the peak density remains nearly constant over a range of driving frequencies - an occurrence observed for low pressure operating conditions i.e. in the collisionless regime. Our present simulations have been carried out for a fixed pressure of 5 mTorr but similar behavior can be expected at higher pressures as long as the device parameters ensure that the electron mean free paths remain equal to or greater than the system length. The constant density regime disappears as soon as the plasma turns highly collisional \cite{Wil_24_2015}. Other factors that might influence the existence of such a parametric domain besides the gap width between the electrodes are the type of gas used, the electrode properties, etc.

\bibliographystyle{unsrt}

\end{document}